\documentclass[a4paper,11pt]{article}        
\pdfoutput=1
\usepackage{jheppub}
	
\usepackage[T1]{fontenc}
\usepackage[utf8]{inputenc}
\usepackage{float}

\title{Phenomenology of the trilinear Higgs coupling at proton-proton colliders}

\author[1]{Magdalena Slawinska,}
\author[1,2]{Wouter van den Wollenberg,}
\author[1]{Bob van Eijk,}
\author[1,3]{Stan Bentvelsen}

\affiliation[1]{Nikhef, Science Park 105, 1098 XG, Amsterdam, The Netherlands}
\affiliation[2]{University of Twente, Drienerlolaan 5, 7522 NB Enschede, The Netherlands}
\affiliation[3]{University of Amsterdam, Spui 21, 1012 WX Amsterdam, The Netherlands}

\abstract{We investigate Higgs pair production at proton-proton colliders, with emphasis on
the gluon fusion channel at the HL-LHC. 
We study the behaviour of the leading order matrix element 
using exact computation of quark loops and infinite quark mass approximation.
We analyse di-Higgs kinematics in search for phase space regions where
the contribution of Higgs self-coupling to SM Higgs pair production is enhanced. 
We discuss how non-SM values of the Higgs trilinear coupling
may affect the kinematics of the Higgs pair.}

\begin{document}
\maketitle
\flushbottom

\section{Introduction} 
Recently, a new scalar boson has been  discovered at the Large Hadron Collider 
at CERN \cite{Aad:2012tfa, Chatrchyan:2012ufa}.
The scalar has a mass of approximately 125 GeV and decays into fermions and gauge bosons 
at a rate consistent with the predictions for a
Standard Model (SM) Higgs~\cite{ATLASHiggsProperties, CMSHiggsProperties}.
If the new particle is indeed the SM Higgs boson, it explains electroweak symmetry breaking
and completes the particle content of the SM.
To understand the dynamics of electroweak symmetry breaking, a detailed measurement of the Higgs potential
is indispensable.
%
The SM Higgs mechanism \cite{Englert:1964et, Higgs:1964pj} 
is responsible for generating the mass of 
electroweak vector bosons via a non-zero vacuum expectation value (VEV) of the Higgs field ($\phi$)
and restores the unitarity of the theory.
The potential of the Higgs field has the form:
\begin{equation}\label{eq:potential}
V(|\phi|^2) = \mu^2|\phi|^2 + \lambda |\phi|^4,
\end{equation}
where $\lambda >0$ and $\mu^2 < 0$. The dependence on $|\phi|^2$ is motivated by gauge invariance 
and the polynomial form 
gives  the simplest expression for a renormalisable potential.
The minimum value of the Higgs potential is the VEV, $v^2 = -\mu^2/\lambda$. 
After spontaneous symmetry breaking, $\phi= (v + H^0)/\sqrt{2}$, 
where  $H^0$ is the excitation from the VEV, 
the Higgs boson acquires mass $m_H^2 = -2\mu^2=2\lambda v^2$, as well as 
cubic and quartic self-interactions:

\begin{equation}
\begin{split}
V(H^0) &= 
2\lambda v^2 \frac{ (H^0)^2}{2} + 6\lambda v \frac{(H^0)^3}{3!} + 6\lambda\frac{(H^0)^4}{4!} 
- \frac{v^4\lambda}{4} \\
&\equiv m_H^2 \frac{(H^0)^2}{2} + \lambda_{3H}\frac{(H^0)^3}{3!}  + \lambda_{4H}\frac{(H^0)^4}{4!} 
- \frac{v^4\lambda}{4}.
\end{split}
\end{equation}

The SM Higgs triple and quartic couplings are uniquely 
defined and read:
\begin{equation}\label{eq:l3Hdef}
\lambda_{3H}= \frac{3 m_H^2}{v}, \quad
\lambda_{4H}= \frac{3 m_H^2}{v^2}.
\end{equation}
In this paper we focus on the triple Higgs coupling $\lambda_{3H}$.
We review the production of two Higgs bosons in a single collision at hadron colliders
and demonstrate that only a fraction of these events is due to processes involving Higgs self-couplings.
To determine the size of signal and background, we revisit the mechanisms of Higgs pair
production. In section~\ref{sec:analytical} we concentrate on production via gluon-gluon fusion and a quark 
loop.
We compare exact and approximate leading order matrix elements. 
As the main contribution to the quark loop stems 
from the top quark,
the exact calculation includes the proper (physical) top quark mass while the approximation takes the limit
 in which the top quark mass becomes infinitely large
 (`effective field theory' - EFT~\cite{Djouadi:1991tka,Spira:1995rr,Dawson:1990zj}).  
Next, in section~\ref{sec:differential}, we calculate cross-sections at the LHC 
and analyse
the di-Higgs kinematics. The emphasis is on kinematical properties that may enhance the terms including 
the trilinear Higgs coupling
versus SM Higgs pair production.
In Section~\ref{sec:background} we discuss potential decay channels for identifying events with Higgs 
pairs at the 
HL-LHC and estimate production cross-sections for Higgs pairs at a 100 TeV hadron collider.
In Section~\ref{sec:BSM} we compare SM cross-sections with those in which the trilinear Higgs coupling 
is modified by beyond SM physics.

\section{Higgs pair production through gluon fusion}\label{sec:analytical}
\begin{figure}
\begin{center}
\includegraphics[width=6cm]{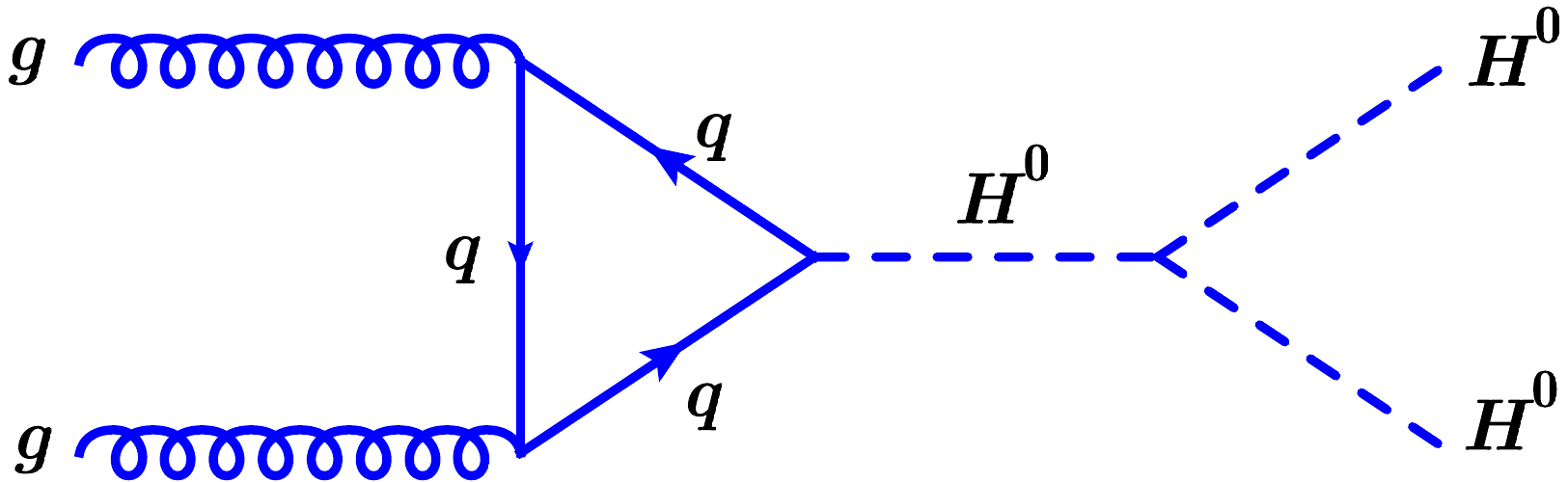}
\hspace{1cm}
\includegraphics[width=6cm]{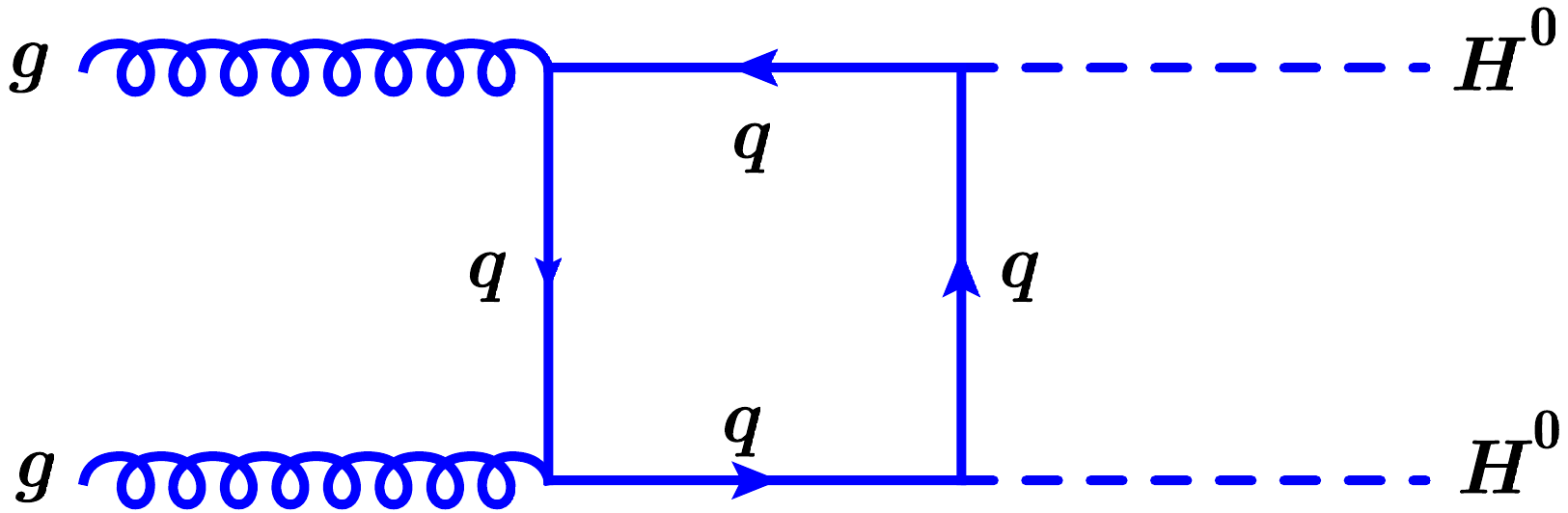}
\caption{Leading order Feynman diagrams for SM Higgs pair production in gluon-gluon fusion.
}
\label{fig:diagrams}
\end{center}
\end{figure}

 \begin{table}
 \begin{center}
\begin{tabular}{|c|r|r|}
      \hline
     Process & Order & $\sigma(pp\to H^0H^0)$ [fb]\\[5pt] 
       \hline
\; & \; & \; \\[0.5pt]
       $gg\rightarrow H^0H^0$ & LO~\cite{Dawson:1998py}	 & 16.5$^{+4.6}_{-3.5}$ \\[5pt]
      (gluon-gluon fusion)  &  NLO~\cite{Dawson:1998py} & 31.9$^{+5.5}_{-4.6}$\\[5pt]
        & NNLO~\cite{deFlorian:2013jea} & 40.2$^{+3.2}_{-3.5}$ \\[5pt]
       \hline
\; & \; & \; \\[0.5pt]
      $qq \rightarrow qqH^0H^0$ & LO~\cite{Dittmaier:2011ti} & 1.81$^{+0.16}_{-0.14}$\\[5pt]
      (vector boson fusion) &  NLO~\cite{Baglio:2012np} & 2.01$^{+0.03}_{-0.02}$\\[5pt]
      \hline      
\; & \; & \; \\[0.5pt]
      $qq\rightarrow W^{\pm}H^0H^0$ & LO~\cite{Dittmaier:2011ti} &0.43$^{+0.005}_{-0.006}$ \\[5pt]
\; & \; & \; \\[5pt]
      (associated production) &  NNLO~\cite{Baglio:2012np} & 0.57$^{+0.0006}_{-0.002}$\\[5pt]
      \hline      
\; & \; & \; \\[0.5pt]
      $qq\rightarrow Z^0H^0H^0$ & LO~\cite{Dittmaier:2011ti} & 0.27$^{+0.004}_{-0.004}$\\[5pt]
\; & \; & \; \\[5pt]
     (associated production) & NNLO~\cite{Baglio:2012np} & 0.42$^{+0.02}_{-0.02}$ \\ [5pt]
     \hline 
      \end{tabular}
      \caption{Dominant cross sections for SM Higgs pair production at the LHC at $\sqrt{s} = $ 14 TeV. 
The errors account for scale uncertainties  only, which in case 
of associated production with W at NNLO are a factor of 10 
smaller than uncertainties due to parton distribution functions.}
      \label{tab:production}
      \end{center}
 \end{table}
In proton-proton collisions, the most important processes contributing to 
events with two Higgs bosons in the final state
are presented in Table~\ref{tab:production}. 
Both leading order (LO) 
and higher order (NLO and NNLO) cross-sections are listed.
The dominant production channel is gluon-gluon fusion, exceeding vector boson fusion 
by a factor of $\sim$20.
The Higgs pair production channels listed in Table~\ref{tab:production} include 
diagrams for both self-coupling and where two Higgses are produced separately. 
Quoted errors reflect scale uncertainties only ($\sqrt{\hat{s}}/2 < \text{scale} < 2\sqrt{\hat{s}}$).
The leading order Feynman diagrams for Higgs pair production in gluon-gluon fusion
are shown in Fig.~\ref{fig:diagrams}. Only the `triangle' diagram contains the trilinear Higgs coupling.
To obtain the di-Higgs cross-section, both diagrams and their interference need to be
evaluated.

The  expression for the  partonic cross-section is given by~\cite{Dawson:1998py, Glover:1987nx}:

\begin{equation} 
\hat{\sigma}^{(LO)}_{gg\to H^0H^0} =
\int d \hat{t}\; \frac{\alpha^2\alpha_S^2}{2^{15}\pi M_W^4}
\left(
  | C_{\triangle} F_{\triangle}+ C_{\square} F_{\square}|^2
     \right),
\end{equation}
where $C_{\triangle}F_{\triangle}$ and $C_{\square}F_{\square}$ correspond to the
individual contributions from the diagrams in Fig.~\ref{fig:diagrams}.
The scale in $\alpha_S$ has been set to the invariant mass of the two incoming partons, $\sqrt{\hat{s}}$.
The Mandelstam variable $\hat{t}$ is defined as:
\begin{equation}\label{eq:hatt}
\hat{t} = -\frac{1}{2} \left[
\hat{s} -2m_H^2 -
\hat{s}\sqrt{1- \frac{4m_H^2}{\hat{s}} } \cos\theta,
\right],
\end{equation}
where $\theta$ is the angle between the two final state Higgs bosons
in the centre of mass frame.
The term in the matrix element squared (MES)
\begin{equation}\label{eq:ME}
 | C_{\triangle} F_{\triangle}+ C_{\square} F_{\square}|^2
\end{equation}
requires the calculation of the form factors  
$F_{\triangle}$ and $F_{\square}$ stemming from the triangle and box loops. 
The coefficients
$C_{\triangle}$ and $C_{\square}$ express the resonance behaviour of the Higgs propagators. 
$F_{\triangle}$ and $F_{\square}$ can be calculated either
exactly or by applying EFT e.g. in the limit where the top quark mass becomes infinite.
To our knowledge the analytical comparison between exact and EFT MES
has never been performed for a light Higgs ($m_H = 125$ GeV).
We investigate the quality of EFT approximation in the following.

The exact formula for $F_{\triangle}$  in eq.~\eqref{eq:ME} is given 
by~\cite{Glover:1987nx} and~\cite{Plehn:1996wb}: 
\begin{equation}
\label{eq:triangle1}
F_{\triangle} = 2\frac{m_q^2}{\hat{s}}
\left [2 + \left(4-\frac{\hat{s}}{m_q^2} \right) m_q^2 C_{ab} \right]=
\tau_q[ 1 + (1- \tau_q) f(\tau_q)],
\end{equation}
where $\tau_q = \frac{4m_q^2}{\hat{s}}$ and $m_q$ is the mass of the fermion in the loop. 
We only consider the top quark as the contribution from bottom and lighter quarks is negligible.
The function $ f(\tau_q)$ stems from  the scalar integral, with
$C_{ab} =-\frac{2}{\hat{s}} f(\tau_q)$ with
\begin{equation}
\label{eq:triangle2}
f(\tau_q) = 
	\begin{cases}
	\arcsin^2\left(\frac{1}{\sqrt{\tau_q}} \right) & \tau_q \geq 1,\\
	-\frac{1}{4}\left[\log\frac{1+\sqrt{1-\tau_q}}{1-\sqrt{1-\tau_q}} - i\pi\right]^2 & \tau_q < 1.
	\end{cases}
\end{equation}
The analytical expression for $F_{\square}$ is too lenghty to present here and
can be found in~\cite{Glover:1987nx, Plehn:1996wb}. 

\begin{figure}[t]
\center
\includegraphics[width=12cm, height=8cm]{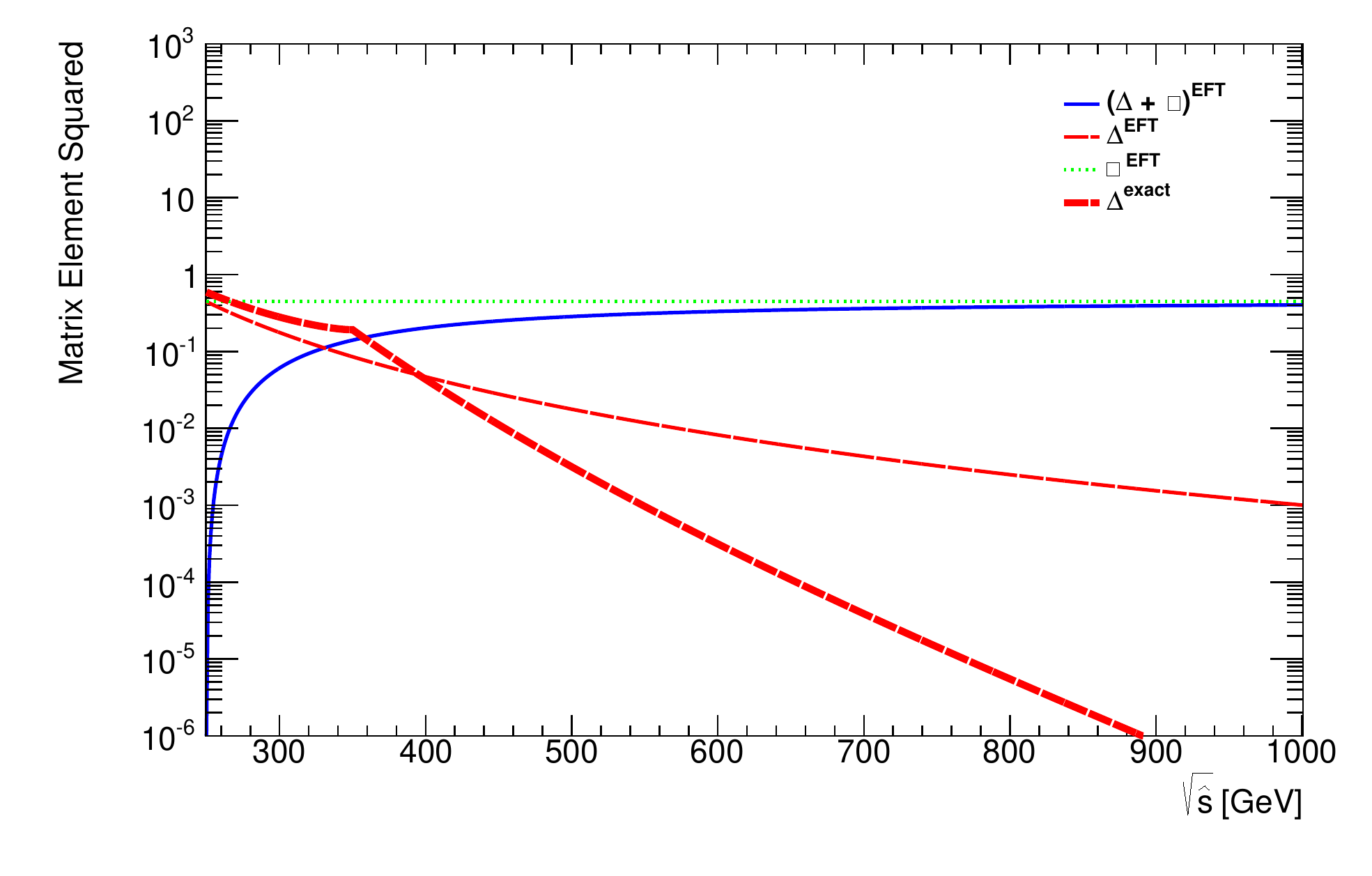}
\caption{Comparison between:
$\triangle^{EFT}\equiv|C_{\triangle}F_{\triangle}^{EFT}|^2$ (red-dashed), $\square^{EFT}\equiv|C_{\square}F_{\square}^{EFT}|^2$ (green-dotted),
$(\triangle + \square)^{EFT}\equiv|C_{\triangle}F_{\triangle}^{EFT} + C_{\square}F_{\square}^{EFT}|^2$ (blue-solid), 
$\triangle^{\text{exact}}\equiv |C_{\triangle}F_{\triangle}|^2$ (thick red-dot-dashed).
}
\label{fig:analytical}
\end{figure}

The Higgs coefficients are:
\begin{equation}\label{eq:Cformfactors}
C_{\triangle} =\frac{\lambda_{3H} v}{\hat{s}-m_H^2}, \;
C_{\square} = 1.
\end{equation}
After series expansion of the first term in eq.~\eqref{eq:triangle2}, $f(\tau_q)$
can be written as  
\begin{equation}
f(\tau_q) = \frac{1}{\tau_q} + \frac{1}{3\tau_q^2} + 
\mathcal{O}\left (\left(\frac{\hat{s}}{4m_q^2} \right)^3\right).
\end{equation}
In the infinite top mass approximation, $\tau_{q}\rightarrow +\infty$ and
$F_{\triangle}$ becomes 
\begin{equation}\label{eq:formfactortriangle}
F_{\triangle}^{EFT} = 2/3,
\end{equation}
while~\cite{ Plehn:1996wb}
\begin{equation}\label{eq:EFTformfactorbox}
F_{\square}^{EFT} = -2/3. 
\end{equation}
The comparison between  EFT and the exact expression for the box MES we leave for future study.
The expression in~\eqref{eq:ME} reduces to:
\begin{equation}\label{eq:MEEFT}
\left|F_{\triangle}^{EFT}C_{\triangle}+ F_{\square}^{EFT}C_{\square}\right|^2=
(2/3)^2
\left (
\frac{\lambda_{3H}^2 v^2}{(\hat{s}-M_H)^2} -   \frac{2\lambda_{3H} v}{\hat{s}-M_H} + 1
\right).
\end{equation}
In Fig.~\ref{fig:analytical} the behaviour
of $|C_{\triangle}F_{\triangle}^{EFT}|^2$ (dashed line), $|C_{\square}F_{\square}^{EFT}|^2$ (dotted line),
$|C_{\triangle}F_{\triangle}^{EFT} + C_{\square}F_{\square}^{EFT}|^2$ (solid line) 
as a function of $\sqrt{\hat{s}}$ is displayed.
In addition, $|C_{\triangle}F_{\triangle}|^2$ (dot-dashed line) is depicted.
The genuine self-coupling contribution is only important for $\sqrt{\hat{s}}$ smaller than 400 GeV.
The box contribution however, dominates over almost the full range for $\sqrt{\hat{s}}$.
Near $\sqrt{\hat{s}} \simeq 2\times m_H$
the  interference between the triangle and box
 leads to sizeable cancellations. 
The figure also demonstrates that there is a large discrepancy between exact and approximate calculations
of the triangle contribution for $\sqrt{\hat{s}} \gg 2 m_q$.
The triangle contribution to the MES contains the intermediate Higgs propagator,
which is probed far off-shell at $\sqrt{\hat{s}} \gg m_H$. 
Therefore, this contribution becomes much smaller than the box, in which this propagator is absent.
For increasing $m_q$ in eqs.~\eqref{eq:triangle1} and \eqref{eq:triangle2}
the agreement between EFT and exact for the triangle improves.

\begin{figure}[t]
\center
\includegraphics[width=12cm, height=8cm]{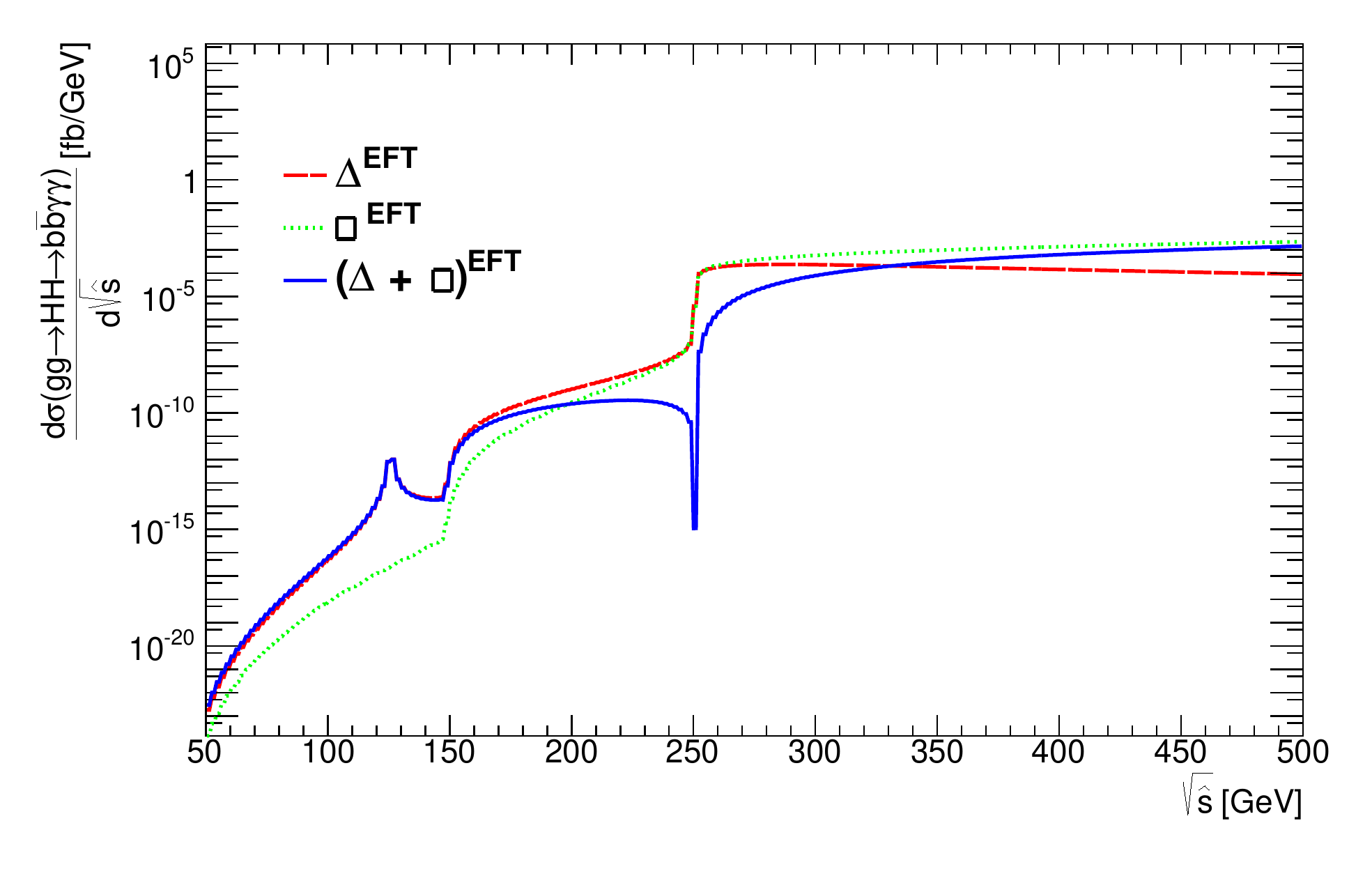}
\caption{The $gg\to H^0H^0 \to b\bar{b}\gamma\gamma$  cross-section as a function of  $\sqrt{\hat{s}}$.
The results are obtained
using  Madgraph5 (EFT)~\cite{Alwall:2014hca}. Red-dashed line -- `triangle' only contribution, green-dotted line -- only `box' contribution,
 Blue-solid line -- full result. }
\label{fig:s-hat}
\end{figure}

We focus on the kinematical region $\sqrt{\hat{s}} < 2m_H$ as we expect that the triangle contribution dominates 
at these low energies. At $\sqrt{\hat{s}} < 2m_H$  the final state Higgses
can no longer be both on-shell and the expression for the MES in eq.~\eqref{eq:ME} is no longer valid.
To cover the range where one or both final state Higgs bosons are off-shell we
force one Higgs to decay into $b\bar{b}$ and the other into $\gamma\gamma$. 
We analyse the behaviour of the partonic cross-section for $gg\rightarrow H^0H^0 \rightarrow b\bar{b}\gamma\gamma$
numerically  with  Madgraph5~\cite{Alwall:2014hca} (EFT). 
The various contributions to this process, as a function of 
$\sqrt{\hat{s}}$,
are shown in  Fig.~\ref{fig:s-hat}.
The choice of decay modes affects the total normalisation
but not the shape of the distributions.
The following cuts are used to ensure that the result is free from phase space singularities: 
\begin{itemize}
\item pseudorapidity for each final state particle: $|\eta| < $ 2.5,
\item transverse momentum of each final state particle: $p_T> $10 GeV,
\item spatial separation: 
$\Delta R(b\bar{b}), \Delta R(b\gamma), \Delta R(\bar{b}\gamma), \Delta R(\gamma\gamma)> 0.4$, 
\end{itemize}
where $\Delta R=\sqrt{(\Delta \phi)^2 + (\Delta \eta)^2}$, 
$\Delta\phi$ is the angle between two particles 
in the plane perpendicular to the incoming gluons
and $\Delta \eta$ the difference in pseudorapidity.
The region below $\sqrt{\hat{s}}=125$ GeV in Fig.~\ref{fig:s-hat} 
corresponds to all Higgs bosons being off-shell,
in the intermediate region $125 \text{ GeV }<\sqrt{\hat{s}}< 250\text{ GeV}$ at least one Higgs is allowed to be on-shell. 
Obviously, above $250$ GeV both favour to be on-shell. Due to the narrow Higgs width both the one and the two Higgs
resonances appear as `kinks'. 
In the vicinity of $\sqrt{\hat{s}}=$125 GeV the rise of the total partonic cross-section is the result of
the large influence of the Higgs propagator, as the triangle dominates.
Close to $\sqrt{\hat{s}}=$250 GeV, on the other hand, the box contribution is equally important.
The partonic cross-section drops due to strong negative interference.
Another  `kink' appears near $\sqrt{\hat{s}} = 145$ GeV. This is the result of the
cut on the transverse momenta of the b-quarks and photons.
In conclusion, only for the region $\sqrt{\hat{s}} < 200$ GeV
the partonic cross-section is dominated by the contribution from the trilinear coupling.
This unfortunately implies that the trilinear Higgs coupling adds only a small fraction to the total cross-section. 

\section{Higgs pair differential cross-section}\label{sec:differential}
\begin{figure}[t]
\center
\includegraphics[width=12cm, height=8cm]{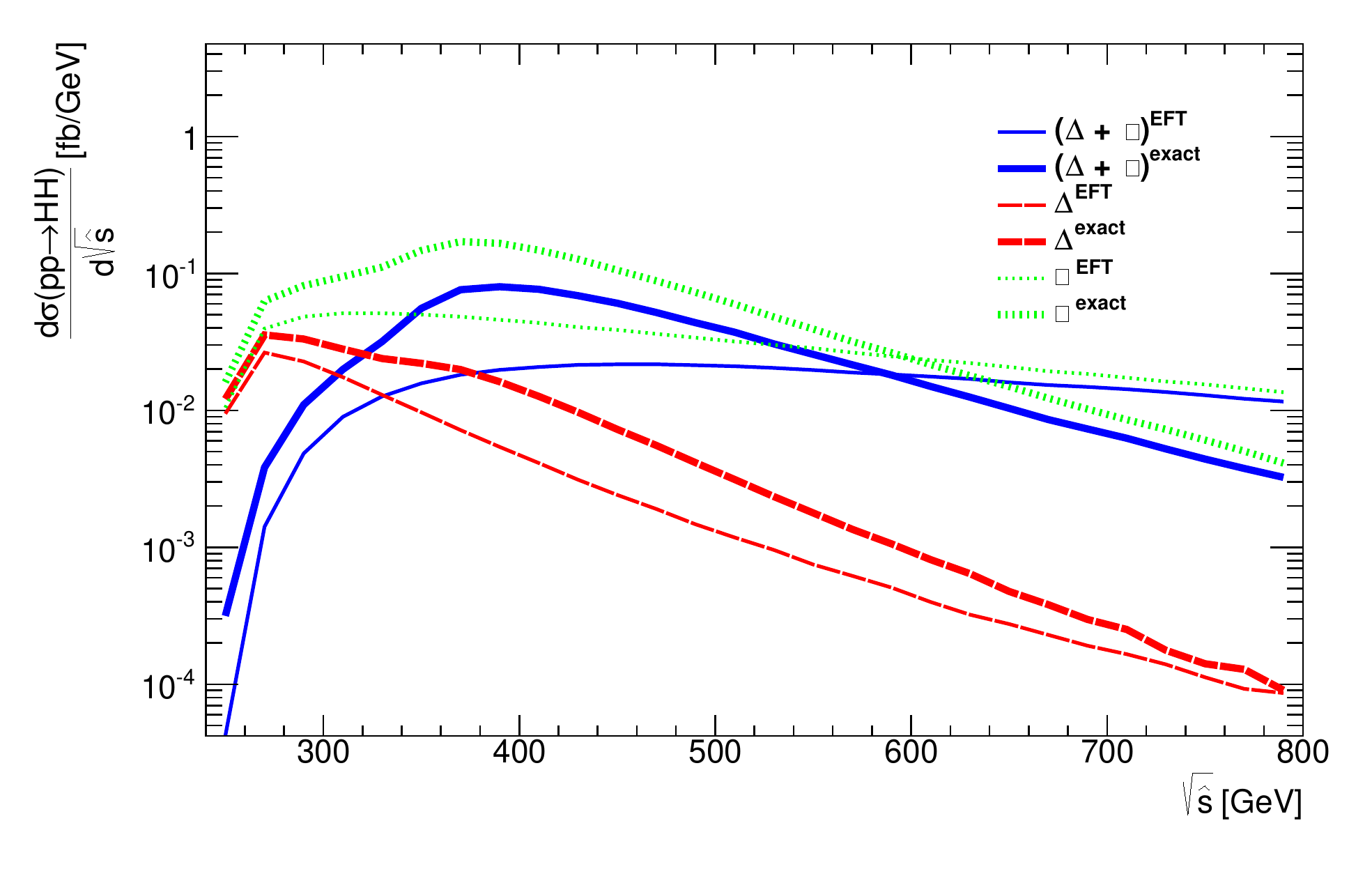}
\caption{The differential Higgs pair production cross-section at leading order for
the triangle, box, and both diagrams (including their interference). 
All contributions are obtained using both EFT and exact calculations (see legend).
}
\label{fig:invmass_boxandboth}
\end{figure}
In the following we convolute MES with gluon density functions (CTEQ6l~\cite{Pumplin:2002vw}) and 
compare the leading order Higgs pair production cross-section $\sigma(pp\to H^0H^0)$
for exact and EFT calculations.
For both we set the factorisation scale equal to the renormalisation scale
($\sqrt{\hat{s}}$).

In Fig.~\ref{fig:invmass_boxandboth} we present two sets of curves. 
One set is obtained using EFT approximation (thin lines), the other set is the result of the  exact approach  (thick lines).
Each set displays the differential cross-section for  the triangle (dashed-dotted line) and  
box (dotted line) diagrams, and their sum (solid line).  
The exact  calculations are obtained using ~\cite{Frederix:2014hta}.
We adapted Madgraph5\footnote{The EFT model in Madgraph5 does not  
    include  the ggHH coupling which,  
    we implemented using Feynman rules given in~\cite{Dawson:1998py}.
    Since each vertex is introduced separately, we obtain individual contributions for the triangle and 
    box diagrams  by forcing either of the couplings ggH=0 or ggHH=0.}~\cite{Alwall:2014hca}  to perform the EFT calculations. 
Despite the steep rise of the gluon density at small values of $x$,
the cross-section below $\sqrt{\hat{s}}= 250$ GeV is negligible. 
At $\sqrt{\hat{s}}=400$ GeV the exact calculation exceeds the EFT approximation for all cases.
We find  a large discrepancy between exact approach and EFT.
The total cross-section at $\sqrt{\hat{s}}< 400$ GeV
is dominated by the interference between triangle and box which, 
leads to large cancellations at $\sqrt{\hat{s}}= 2\times m_H = 250$ GeV. Above this threshold 
the box dominates.
The largest difference between  exact and EFT appears for the 
the box only case. As a consequence, the total cross-section in EFT is underestimated
at low and overestimated at high $\sqrt{\hat{s}}$.
The triangle contribution in both  exact and EFT  is  similar.  
EFT does not reproduce the kink at $2\times m_q$ (which is the result of using
the approximate form factor of eq.~\eqref{eq:formfactortriangle} instead of the one in
eqs.~\eqref{eq:triangle1}  and \eqref{eq:triangle2}, see Fig.~\ref{fig:analytical}). 
\begin{figure}[t]
\center
\includegraphics[width=6cm, height=4cm]{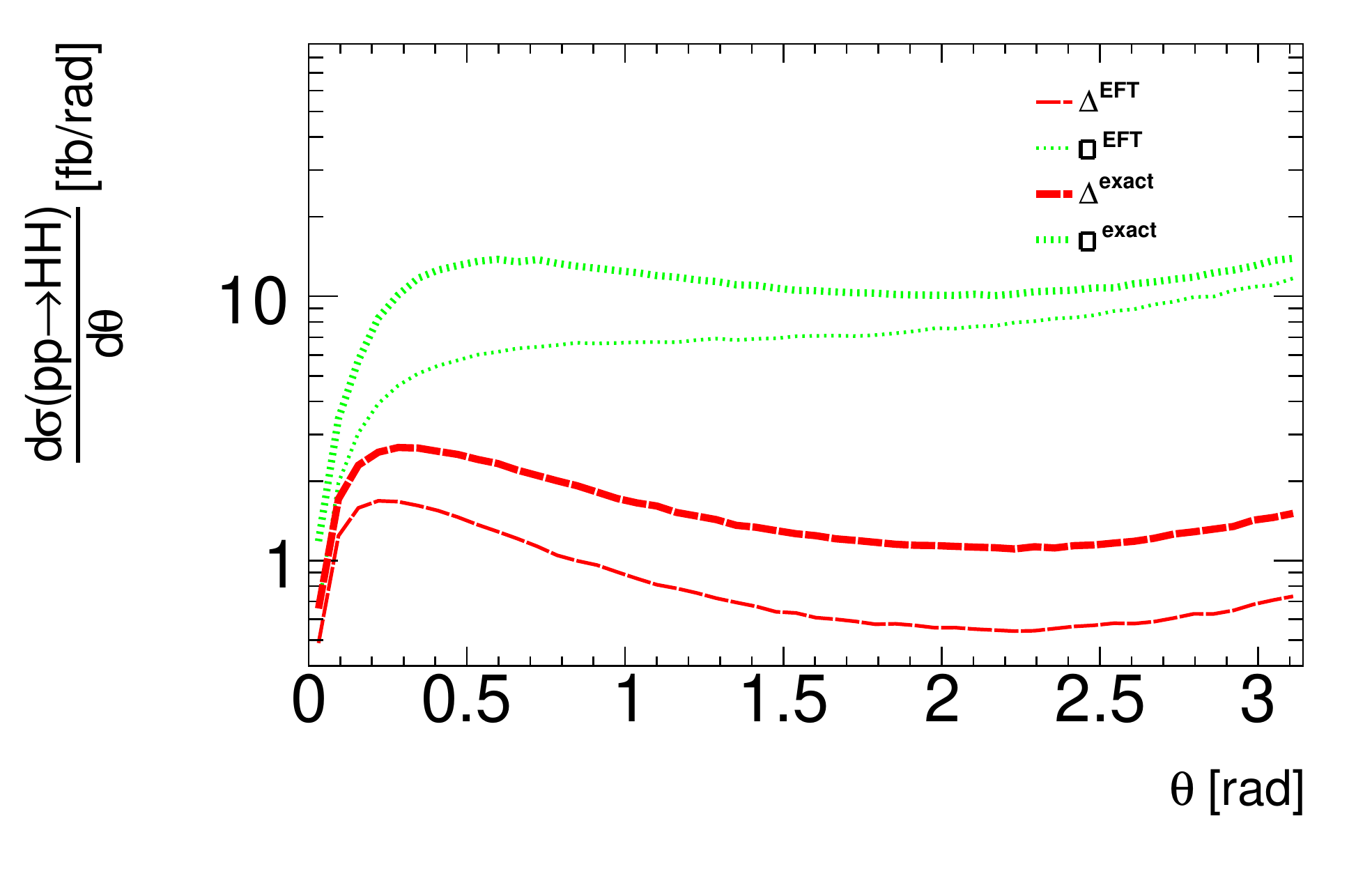}
\includegraphics[width=6cm, height=4cm]{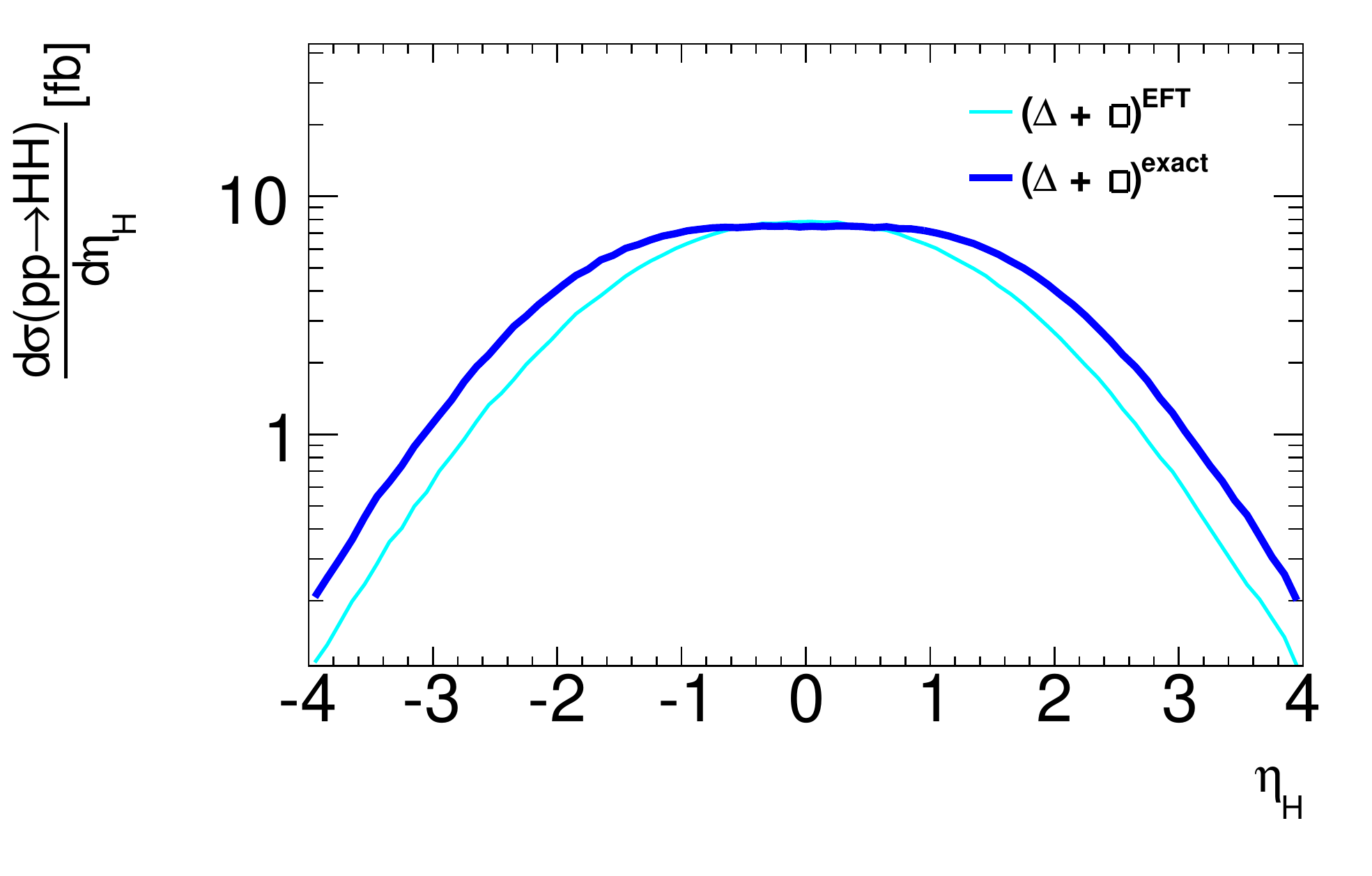}
\caption{Left plot: comparison between the distributions of the opening angle between the two Higgs momenta 
in EFT (thin lines) and exact calculation (thick lines) 
for the triangle (red lines) and box (green lines).
Right plot: 
Comparison between the inclusive Higgs pseudorapidity distributions 
for EFT (thin light line) and exact (thick dark line) calculation.}
\label{fig:anglehiggses}
\end{figure}
In the left plot in Fig.~\ref{fig:anglehiggses} we compare 
the angle between the two Higgses in the laboratory frame
(exact-- thick lines,  EFT--  thin lines)
for the box (dotted) and triangle (dashed-dotted) contributions.
For the triangle, both approaches give similar shapes.
For the box, the difference between exact and EFT calculations 
decreases when the Higgses get more back-to-back. 
The exact calculations show a larger preference for the two Higgses to have a small opening angle.
This effect is less pronounced  in the EFT calculations
and can be explained by analysing the right graph in Fig.~\ref{fig:anglehiggses}.
Here, the rapidity of each Higgs is
presented. Loop calculations result in a broader distribution
which, is caused by larger differences between the Bjorken $x$ of 
the colliding gluons. As a result, the di-Higgs system receives a larger longitudinal 
boost despite lower mean value of $\sqrt{\hat{s}}$.
In conclusion; EFT calculations do not represent the kinematics of Higgs pair production. This is due to   
oversimplification of the  -- dominant -- box contribution.

\section{Disentangling the signal from irreducible background}\label{sec:background}
The discussion in the previous section demonstrates that the kinematical region in which
Higgs pair production cross-section is
most sensitive to the self-coupling contribution, is where $\sqrt{\hat{s}}<400$ GeV.
Next, we will examine how this energy dependence 
affects kinematical properties of the decay products of both Higgses.
We exclude EFT calculations of the signal and study only exact distributions.

In  Table~\ref{tab:yields} the event yields for 
$b\bar{b}b\bar{b}$, $b\bar{b}W^+W^-$, $b\bar{b}Z^0Z^0$, $b\bar{b}\gamma\gamma$, and $4\gamma$ 
final states are given.
We compare three variants:
8 TeV with an integrated luminosity 
of 20 fb$^{-1}$ (as collected by the ATLAS and CMS experiments during the LHC Run I),  
High Luminosity LHC with 3000 fb$^{-1}$ at 14 TeV, and 
 3000 fb$^{-1}$ at a 100 TeV hadron collider.
The numbers are based on Higgs pair production cross-sections 
at LO and NLO~\cite{Dawson:1998py} taking into account the proper Higgs branching ratios~\cite{Dittmaier:2011ti}.
Except for both Higgses decaying into bottom quarks,
the choice of these channels is motivated by the observation of the (single) Higgs
decay into $\gamma\gamma$, $Z^0Z^0$ and $W^+W^-$ by the ATLAS and CMS collaborations~\cite{ATLASHiggsProperties, CMSHiggsProperties}.  
The  $4\gamma$ channel 
suffers from lack of statistics and will therefore be ignored in our analyses.
One might argue, however, that this final state poses a real
 challenge at a 100 TeV hadron collider (21 events).
To provide sufficient statistics at the HL-LHC, at least one  Higgs decay should have a large branching ratio,
for instance $b\bar{b}$ or $\tau^+\tau^-$. 
The $b\bar{b}b\bar{b}$, $b\bar{b}W^+W^-$ and $b\bar{b}Z^0Z^0$  channels suffer from large backgrounds. 
The $b\bar{b}\gamma\gamma$ final state compromises between reasonable statistics and 
a relatively clean experimental signature. 
We examine whether this signal can be distinguished from irreducible $b\bar{b}\gamma\gamma$  backgrounds
resulting from QCD, QED and single Higgs production ($Z^0H(\gamma\gamma$), $b\bar{b}H(\gamma\gamma)$).

We study $pp\to HH \to b\bar{b}\gamma\gamma$ neglecting
initial and final state radiation. Moreover, we do  not take into account
possible dilution effects due to  reconstruction inefficiencies or mis-reconstruction. 
Thus we exclude other (reducible) backgrounds containing for instance 
two photons and two light quark jets,
top anti-top pairs,
top anti-top pairs with single photons, and
top anti-top pairs with Higgses decaying into two photons.
\begin{table}
 \begin{tabular}{ | l | r | r | r | r | r  |}
\hline
HH final state & Br. Rat. &    8 TeV(20 fb$^{-1}$) & 14 TeV (3000 fb$^{-1}$)  & 100 TeV (3000 fb$^{-1}$)\\
 & & LO (NLO)&LO (NLO) & LO (NLO) \\
\hline
&&&&\\
$b\bar{b}b\bar{b}$ & 32.5\% & 23 (60) & 16$\times 10^3$  (33$\times 10^3$)  & 0.85$\times 10^6$ (1.3$\times 10^6$) \\[7pt]
\hline
&&&&\\
$b\bar{b}W^+W^-$ & 23.9\%& 17 (44) & 12$\times 10^3$ (24$\times 10^3$) & 0.63$\times 10^6$ (0.97$\times 10^6$) \\[7pt] 
\hline
&&&&\\
$b\bar{b}Z^0Z^0$ & 3.0\%& 2.2 (5.5) & 1.5$\times 10^3$ (3$\times 10^3$) & 0.08$\times 10^6$  (0.12$\times 10^6$) \\[5pt]
\hline
&&&&\\
$b\bar{b}\gamma \gamma$ & 0.26\%& 0.19 (0.48) & 128 (264) & 6800 (10500)\\[5pt]
\hline
&&&&\\
$\gamma \gamma\gamma \gamma$ & 0.001\%& 0 (0) & 0.25 (0.53)& 14 (21)\\[5pt]
\hline
 \end{tabular}
\caption{Expected event yields for di-Higgs production form gluon fusion and different 
di-Higgs decay channels, based on the LO~\cite{Frederix:2014hta} 
 (NLO \cite{Dawson:1998py}) Higgs pair production 
cross section: 3.58 (9.22) fb, 16.23 (33.86) fb, and 877 (1350) fb at 8, 14, and 100 TeV, respectively.}
\label{tab:yields}
\end{table}
There have been several phenomenological studies on the $b\bar{b}\gamma\gamma$ channel, see
for instance~\cite{Baur:2003gp, Baglio:2012np, Dolan:2012rv}.
We take a different approach and analyse
the  box and  triangle contributions 
separately in order to
find  kinematical regions in which the sensitivity to the 
trilinear coupling is largest. We also discuss the feasibility of
separating the genuine self-coupling from the irreducible background.
We apply the following cuts to our $b\bar{b}\gamma\gamma$ samples~\cite{Baur:2003gp}: 
\begin{equation}\label{eq:cuts}
\begin{split}
& p_T(b) > 45 \text{ GeV}, \; |\eta(b)| < 2.5, \; \Delta R(b,b)>0.4,\; \\
& p_T(\gamma) > 20 \text{ GeV}, \; |\eta(\gamma)| < 2.5, \; \Delta R(\gamma,\gamma)>0.4,
\end{split}
\end{equation}
and
\begin{equation}\label{eq:invmasscuts}
 |m_{bb}-m_{H}|< 20 \text{ GeV}, 
 |m_{\gamma\gamma}-m_{H}|< 2.3 \text{ GeV}.
\end{equation}
The cuts on transverse momentum and rapidity are motivated by 
trigger capabilities and detector coverage. 
The rather tight cuts on invariant masses
increase the signal to background ratio. 
\begin{figure}
\center
\includegraphics[width=12cm, height=8cm]{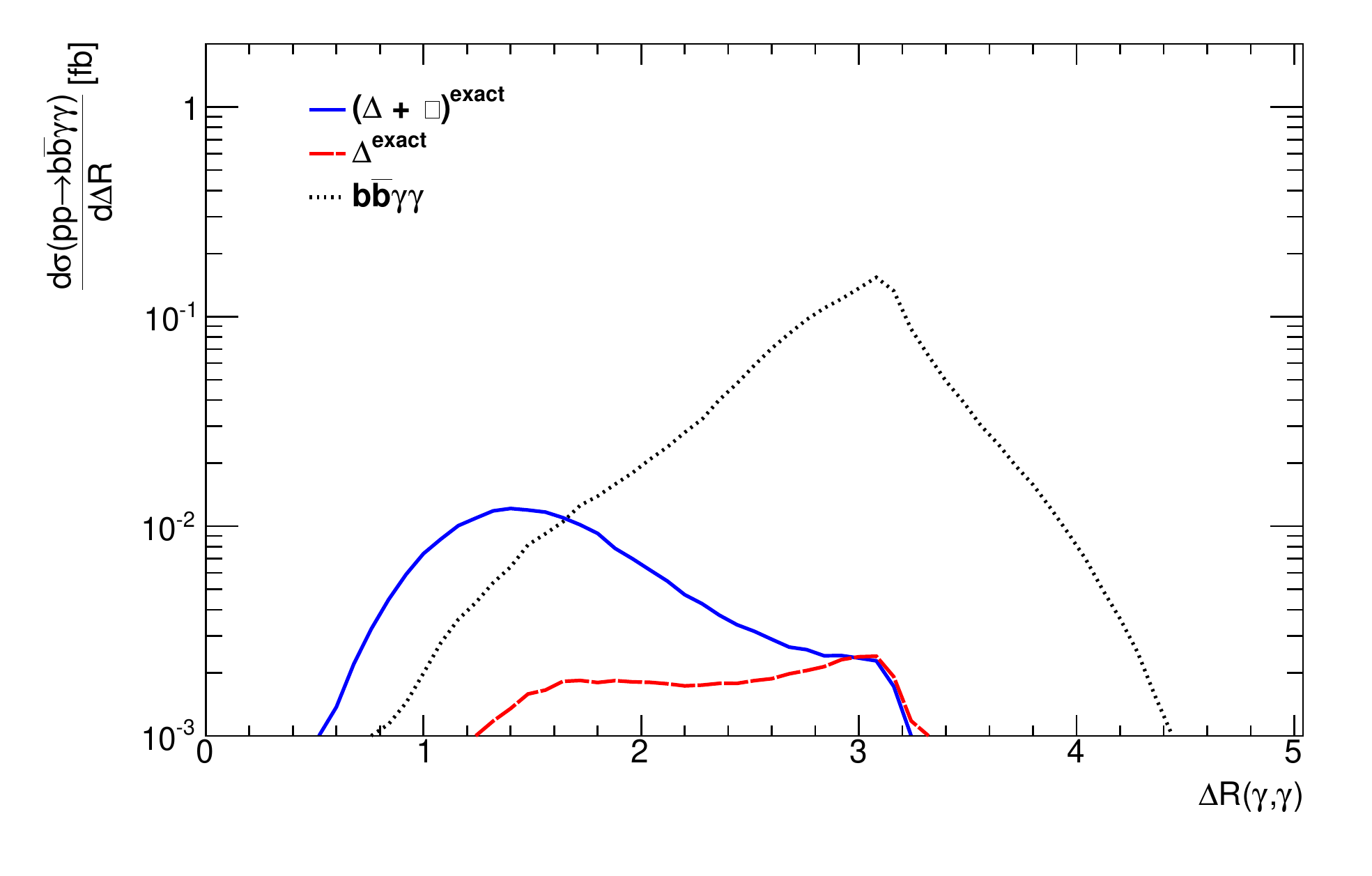} 
\includegraphics[width=12cm, height=8cm]{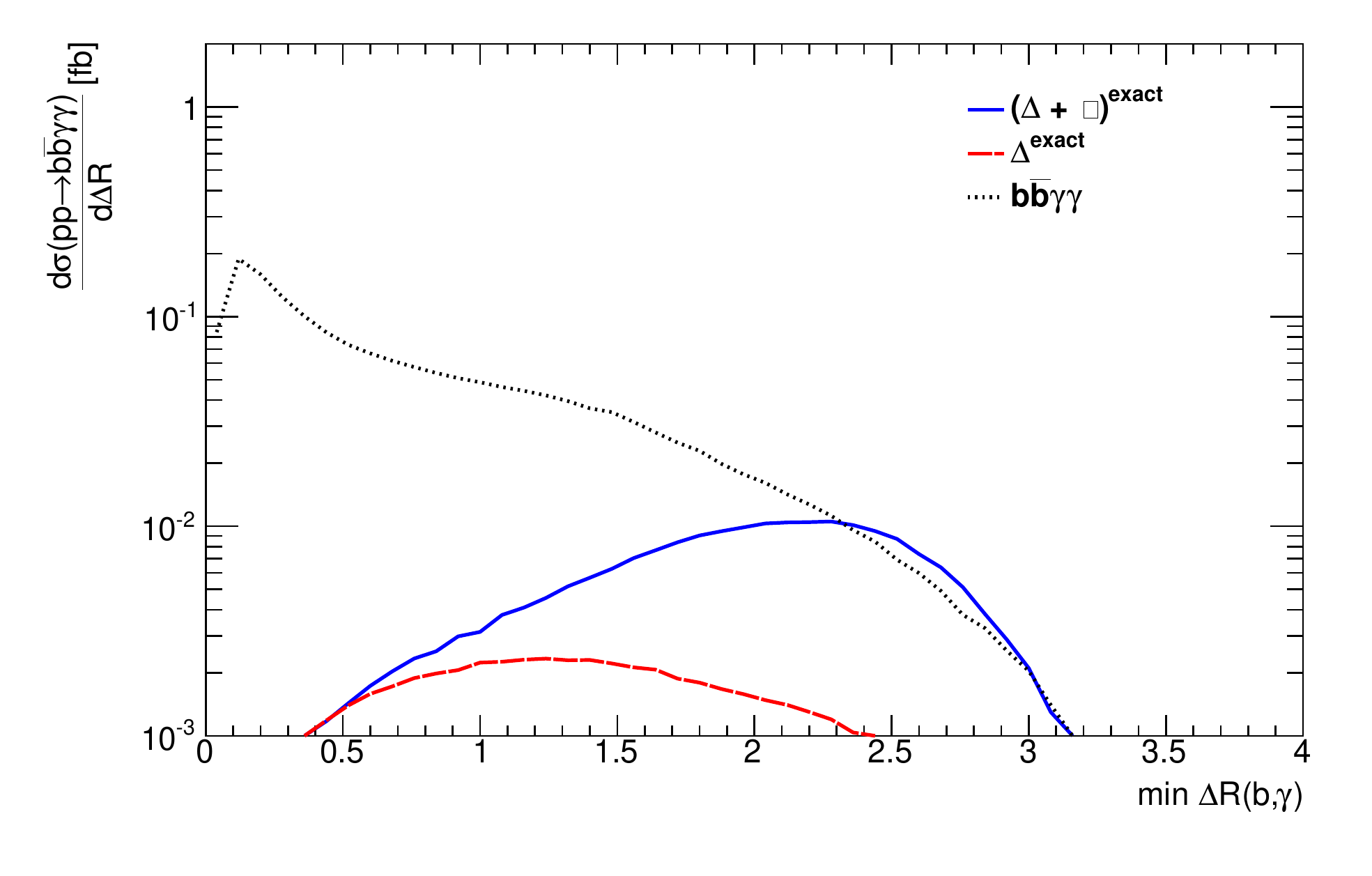}  
\caption{The distributions of $\Delta R(\gamma, \gamma)$ (upper plot) and $\Delta R(b, \gamma)_{\min}$ (lower plot)
for the SM Higgs pair production (blue-solid line), the triangle contribution separately (red-dashed line) and the irreducible $b\bar{b}\gamma\gamma$
background (black-dotted line).}
\label{fig:dRSM}
\end{figure}

In Fig.~\ref{fig:dRSM} the $\Delta R$ separation between the two photons (upper plot) and 
between the photon and the $b$ ($\bar{b}$) quark  that are closest in phase space (lower plot) are displayed. 
The individual distributions for the full cross-section (solid line), 
the genuine self-coupling  contribution (dashed line),  and
$b\bar{b}\gamma\gamma$ background (dotted line) are presented. 
The background processes containing single Higgses are calculated using EFT~\cite{Alwall:2014hca}.
The distribution of $\Delta R$ between photons for the Higgs pair reaches a maximum at $\Delta R(\gamma, \gamma) \simeq 1.5$ due to the large boost 
of the di-Higgs system. It is well separated from the background, in which the two photons 
do not stem from the same parent.
The kinematical properties of the triangle and triangle+box samples are different due to the domination of the box.
The former favours $\Delta R(\gamma\gamma)\simeq 3$ while the latter has a maximum for small values of $\Delta R$. 
The minimum separation between a photon and a $b$ ($\bar{b}$) quark in the background sample is small as
most photons are emitted from the quarks and hence prefer to be collinear with their parents.
In  our two di-Higgs samples on the other hand, these particles
are the Higgs decay products and are much more separated.
The relatively large contribution from the box increases this separation.

Fig.~\ref{fig:dRSM} leads to the conclusion that it is extremally challenging to isolate the genuine
self-coupling contribution from the irreducible background. Different shapes of the overall SM di-Higgs production 
and the trilinear coupling contribution suggest that the strategy optimised to isolate the former might not be
best for enhancing the sensitivity to the latter. We will discuss this issue more quantitatively in the following section.

\section{Non Standard Model values of $\lambda_{3H}$}\label{sec:BSM}
If the magnitude of the Higgs self-coupling is not in accordance with its SM value,
the electroweak symmetry breaking is not, or only in part, a result of the SM Higgs potential 
of eq.~\eqref{eq:potential}. As a consequence, $m_H$, VEV and $\lambda_{3H}$ in eq.~\eqref{eq:l3Hdef}
are decoupled and the observed 125 GeV resonance is not the SM Higgs. 
It may instead be a member of a more extended sector (see for instance~\cite{Georgi:1985nv}) or
be composite and strongly interacting~\cite{Giudice:2007fh,Grober:2010yv}.

\begin{figure}[ht!]
\begin{center}
\includegraphics[width=7.5cm, height=5cm]{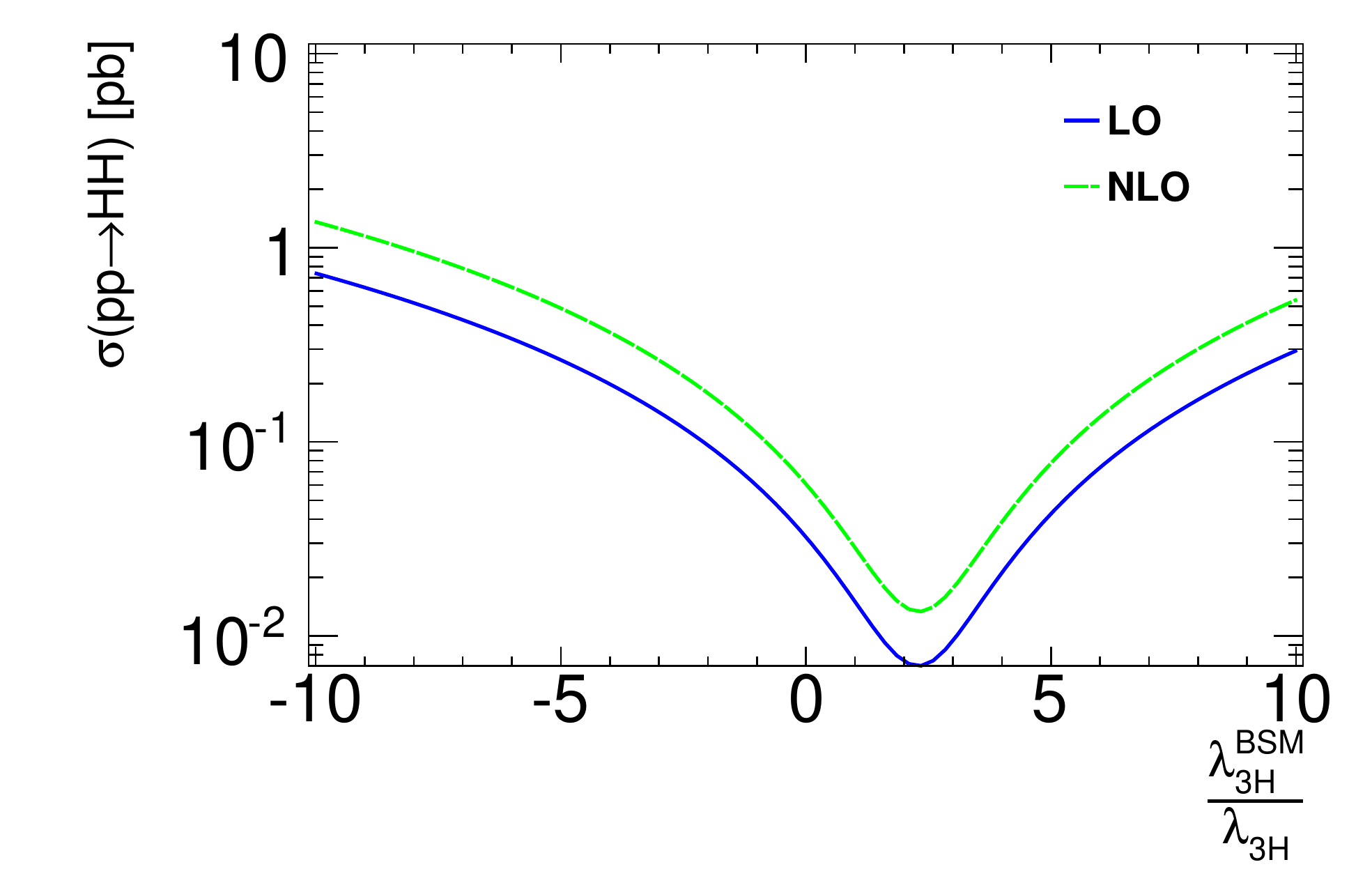}
\caption{Di-Higgs cross-sections at $\sqrt{s}=$ 14 TeV for
different values of $\lambda_{3H}^{BSM}$,
obtained using~\cite{Plehn:1996wb, Dawson:1998py}.}\label{fig:sigmaoflambda}
\end{center}
\end{figure}
\begin{figure}[h]
\begin{center} 
\includegraphics[width=7.5cm, height=5cm]{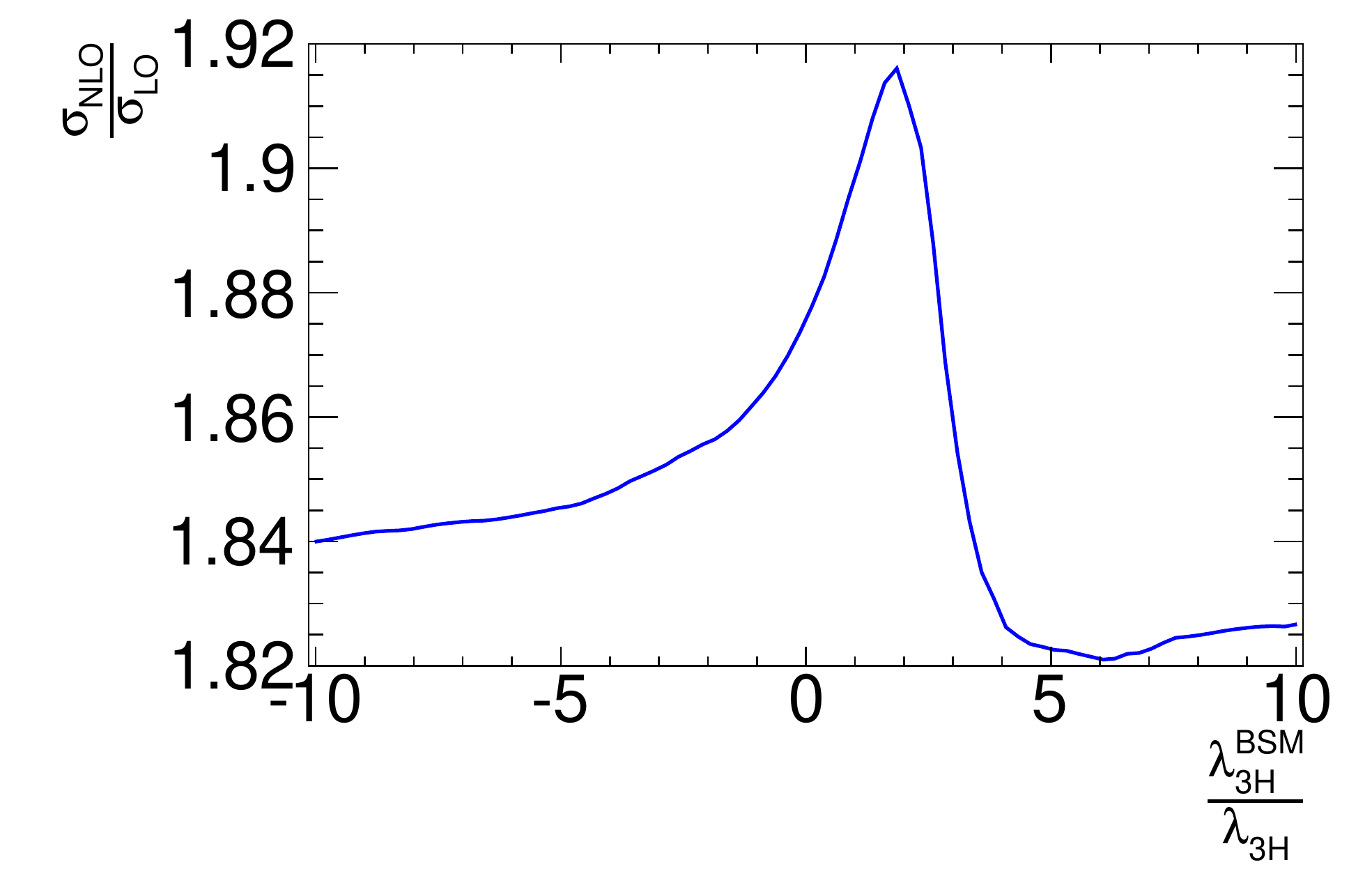}
\includegraphics[width=7.5cm, height= 5cm]{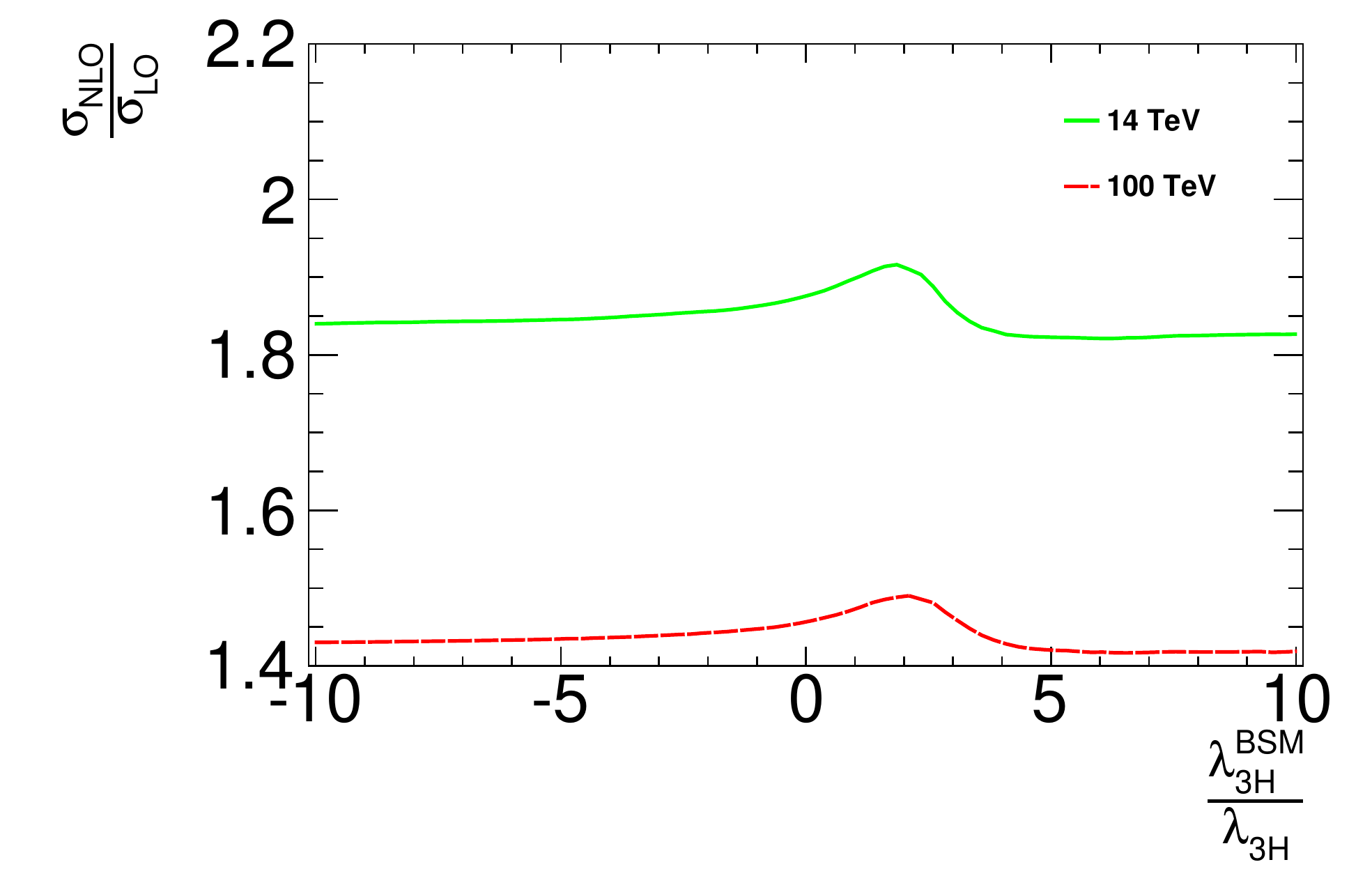}
\end{center}
\caption{NLO K factors as a function of $\lambda_{3H}^{BSM}$ from HPAIR with 
CTEQ6~\cite{Kretzer:2003it} at $\sqrt{s} = 14$ TeV (left plot) and for
$\sqrt{s} = 14$ and $100$ TeV (right plot).}\label{fig:Kfactor} 
\end{figure}

We study how the kinematics in Higgs pair production 
changes assuming the trilinear Higgs coupling is a free parameter.
We do not focus on any model in particular and therefore do not modify any other SM parameter.
In the following $\lambda_{3H}^{BSM}$ denotes the BSM value.
The triangle contribution to the cross-section changes quadratically with $\lambda_{3H}^{BSM}$,
while the triangle-box interference shows a linear dependence.
For  $\lambda_{3H}^{BSM}<0$ the interference term flips sign and the cross-section becomes
larger than for corresponding positive $\lambda_{3H}^{BSM}$.
This behaviour is depicted in Fig.~\ref{fig:sigmaoflambda} showing the 
di-Higgs cross-section at $\sqrt{s} = 14$ TeV
at LO (solid line) and NLO (dashed line)\footnote{
   At NLO the Born term was computed with
   exact quark loop and the QCD corrections were included in EFT approximation. 
   Both LO and NLO were obtained with~\cite{Dawson:1998py}.
   }
 as a function of 
$\lambda_{3H}^{BSM}/\lambda_{3H}$.
Note that for $\lambda_{3H}^{BSM}=0$  only the box contributes and the cross-section  
is larger than for $\lambda_{3H}^{BSM}=\lambda_{3H}$. 
To quantitatively determine the size of the individual contributions we compare LO cross-sections for
$\lambda_{3H}^{BSM}= \lambda_{3H}, 0 \text{ and} -\lambda_{3H}$.
We find that  $\sigma^{\square} \simeq 35$ fb, $\sigma^{int} \simeq -25$ fb, $\sigma^{\triangle}\simeq 5$ fb, while
the total SM cross-section $\simeq 17$ fb.
Due to the large box contribution
the sensitivity of the total cross-section to $\lambda_{3H}^{BSM}$
is small\footnote{The sensitivity of different di-Higgs production channels
to different values of $\lambda_{3H}^{BSM}$ can be found in ref.~\cite{Baglio:2012np}.}.
Fig.~\ref{fig:sigmaoflambda} shows large differences between LO and NLO.
The NLO K-factor ($\equiv\sigma^{NLO}/\sigma^{LO}$)  is displayed in Fig.~\ref{fig:Kfactor} (left)
as a function of $\lambda_{3H}^{BSM}$.
For $\lambda_{3H}^{BSM}=\lambda_{3H}$ the K-factor becomes rather large $\simeq$1.92~\footnote{This value is larger than 
quoted in~\cite{Dawson:1998py} (1.92 instead of 1.86) 
as we used a more recent set of PDFs.} (see Table~\ref{tab:yields}). 
It slightly dependends on $\lambda_{3H}^{BSM}$ due to different QCD corrections 
to box and triangle contributions.
In the right plot in Fig.~\ref{fig:Kfactor} 
the K-factors at $\sqrt{s}=$ 14 TeV and 100 TeV are compared. 
As expected, the role of NLO corrections is smaller at higher $\sqrt{s}$.
The absolute value of the di-Higgs cross-section at 100 TeV is over a factor of 100 larger than at 14 TeV (see Table~\ref{tab:yields}).
The sensitivity to different values of $\lambda_{3H}^{BSM}$ is, however, smaller.

\begin{figure}
\center
\includegraphics[width=12cm, height=8cm]
{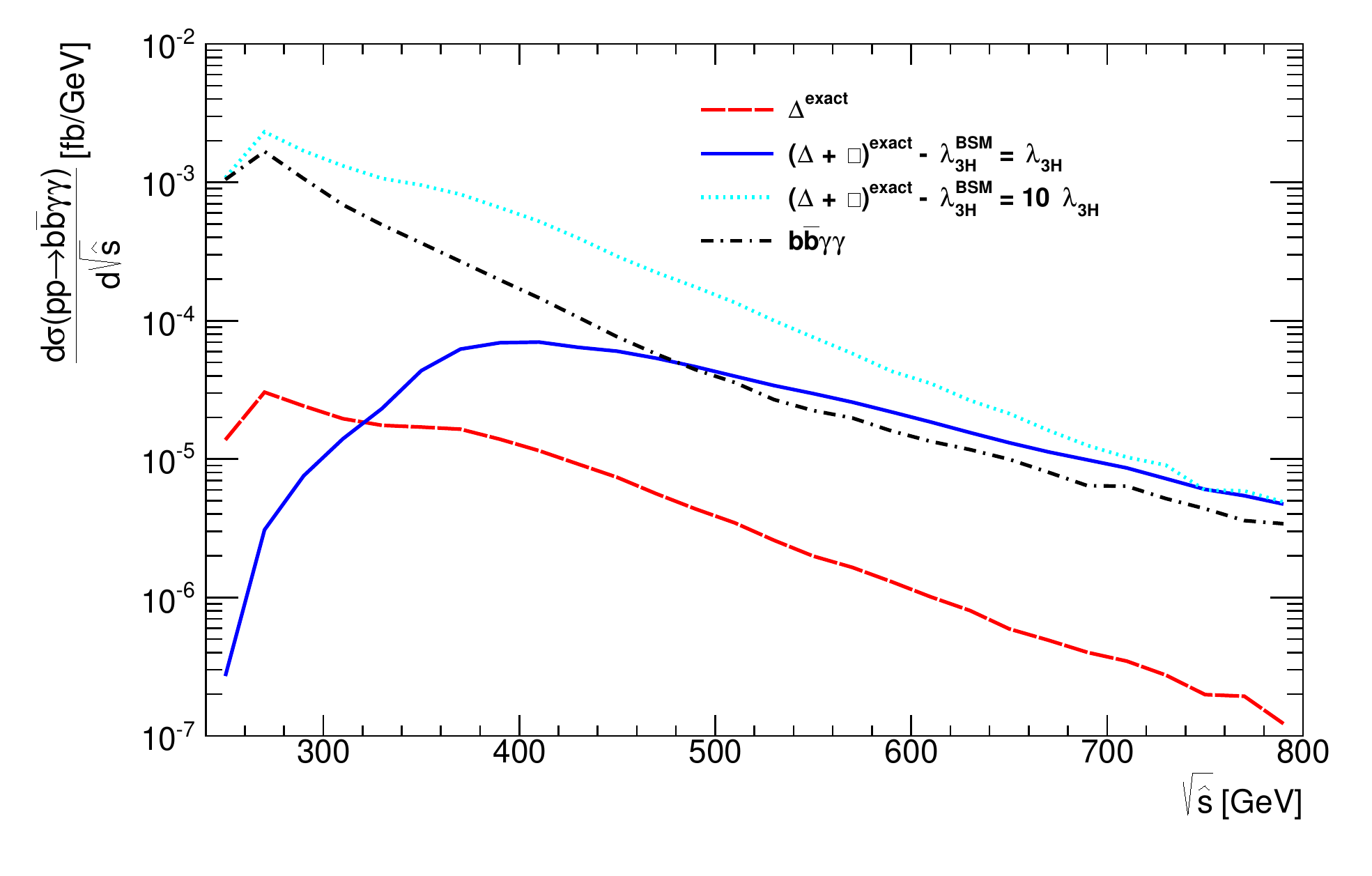}
\caption{The differential di-Higgs production cross-section for $\lambda_{3H}^{BSM} = \lambda_{3H}$ 
(dark blue-dashed line), $\lambda_{3H}^{BSM} = 10\lambda_{3H}$ (light blue-dotted line) and 
genuine self-interactions (red-solid line) and $b\bar{b}\gamma\gamma$ background differential cross-section (black dash-dotted line).
The cuts of eq.~\eqref{eq:cuts} and~\eqref{eq:invmasscuts} are applied.}
\label{fig:invmass}
\end{figure}

Fig.~\ref{fig:invmass}  presents the  differential cross-sections for the processes 
$pp\to HH \to b\bar{b}\gamma\gamma$ and $pp\to b\bar{b}\gamma\gamma$ after applying the cuts of eqs.~\eqref{eq:cuts} and~\eqref{eq:invmasscuts}. 
The SM Higgs pair production is displayed as solid line, the genuine self-coupling as dashed,
BSM Higgs pair production with  $\lambda_{3H}^{BSM}=10 \lambda_{3H}$ as dotted,
and irreducible $b\bar{b}\gamma\gamma$ background as dashed-dotted line.
We focus on BSM with $\lambda_{3H}^{BSM} = 10\lambda_{3H}$. 
The BSM cross-section is approximately  60 times larger than that of the pure triangle  
and two times larger than the irreducible background,
reaches the  maximum at $\sqrt{\hat{s}}= 270$ GeV
and  decreases exponentially for large $\sqrt{\hat{s}}$. 
Unlike the SM cross-section at $\sqrt{\hat{s}}< 400$ GeV, 
at large values of $\lambda_{3H}^{BSM}$ the cancelations due to negative interference are negligible.
On the other hand, at $\sqrt{\hat{s}}> 750$ GeV 
SM and BSM cross-sections become about equal as all sensitivity to Higgs trilinear coupling is lost.

\begin{figure}
\center
\includegraphics[width=12cm, height=8cm]{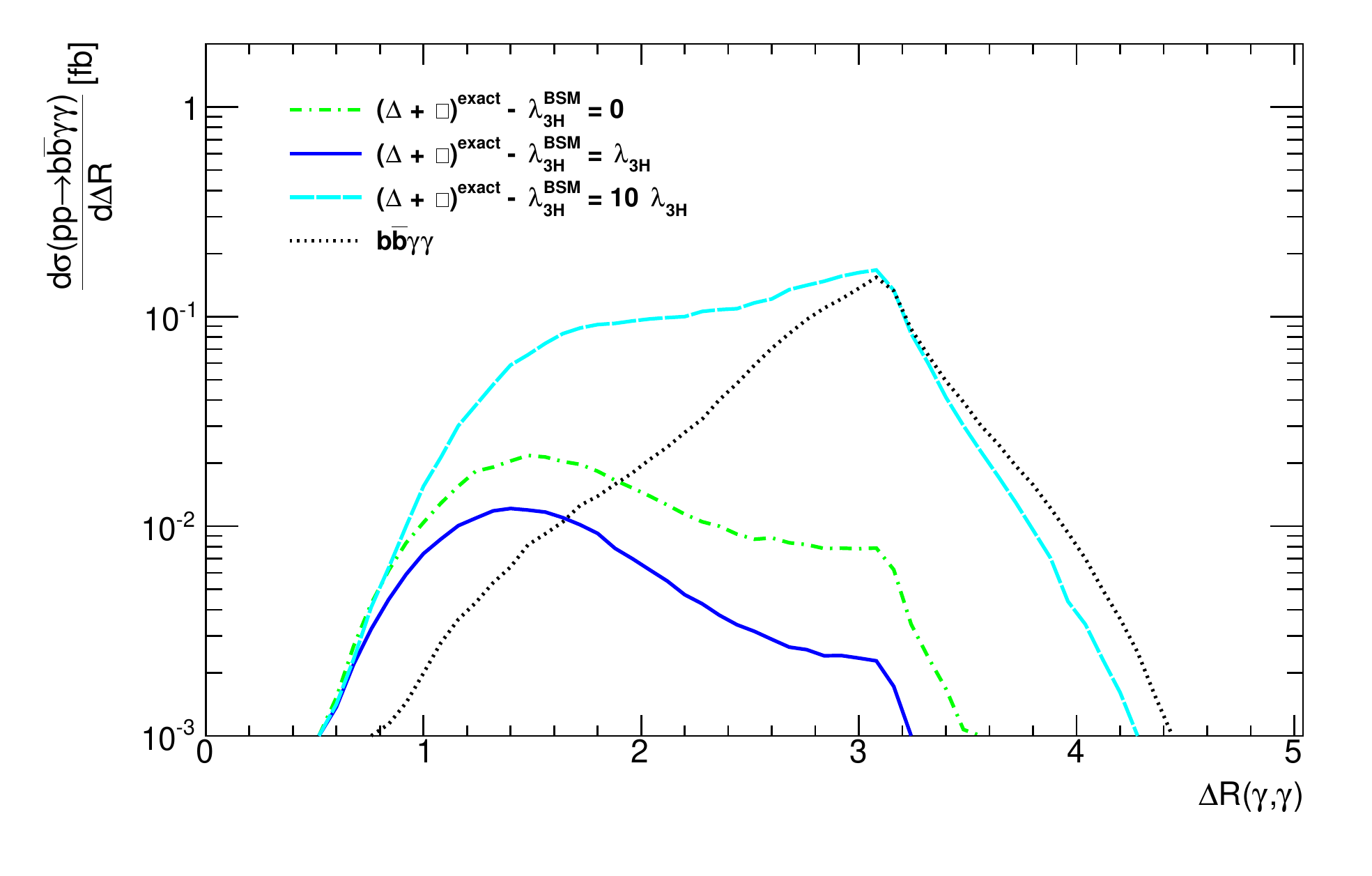} 
\includegraphics[width=12cm, height=8cm]{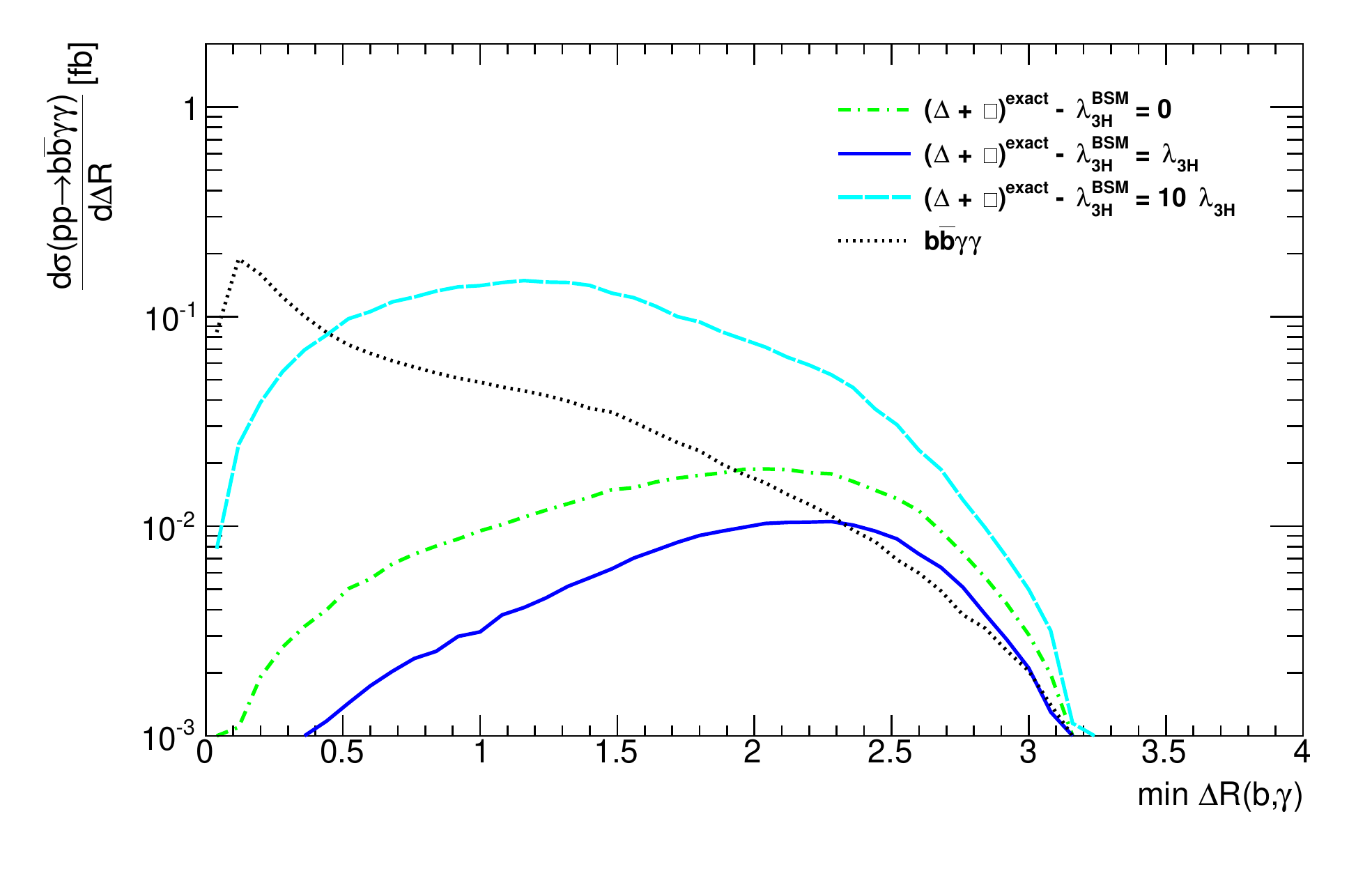}  
\caption{$\Delta R(\gamma, \gamma)$ (upper plot) and $\Delta R(b, \gamma)_{\min}$ (lower plot)
for SM  di-Higgs production (dark blue-solid line), BSM with $\lambda_{3H}^{BSM} = 10 \lambda_{3H}$ 
(dark blue-dashed line), BSM with $\lambda_{3H}^{BSM} = 0$ (green dash-dotted line) and
irreducible background (black-dotted line).
The cuts of eqs.~\eqref{eq:cuts} and \eqref{eq:invmasscuts} have been applied.}
\label{fig:dR_BSM}
\end{figure}

In Fig.~\ref{fig:dR_BSM} we compare $\Delta R$ distributions for the SM and
BSM Higgs pair production mechanisms, 
and for the irreducible background.
The dashed line corresponds to 
$\lambda_{3H}^{BSM} = 10 \lambda_{3H}$,
 dash-dotted line to $\lambda_{3H}^{BSM}=0$,
solid line to the SM  
and dotted line to the $b\bar{b}\gamma\gamma$ background.
As the triangle dominates at small $\sqrt{\hat{s}}$ 
mean separation between the two photons (upper plot) is larger for $\lambda_{3H}^{BSM} = 10 \lambda_{3H}$ than for
$\lambda_{3H}=0$.  
The shape of the pure box sample resembles the SM and its larger cross-section is due to 
the absence of negative interference.
The minimum separation $\Delta R(b,\gamma)$ (upper plot) is on average smaller for $\lambda_{3H}^{BSM} = 10 \lambda_{3H}$
than for pure box.  The larger cross-section for $\lambda_{3H}^{BSM} = 10 \lambda_{3H}$
with $\Delta R(b,\gamma)> 0.4$
on the other hand, separates the triangle from the kinematically similar background.
For $\lambda_{3H}=0$ the signal exceeds the background for $\Delta R(b,\gamma)>2$. 
We proceed with three variants of $\Delta R$ cuts:
\begin{subequations}
\begin{equation}\label{eq:cuts4}
\Delta R(\gamma, \gamma)<2,\quad \Delta R(b, \gamma)>1.0;
\end{equation}
\begin{equation}\label{eq:cuts4a}
\Delta R(\gamma, \gamma)<2.2,\quad \Delta R(b, \gamma)>0.4;
\end{equation}
\begin{equation}\label{eq:cuts4b}
\text{no cut on }\Delta R(\gamma, \gamma),\quad  \Delta R(b, \gamma)>0.4;
\end{equation}
\end{subequations}
with the aim to improve the separation as compared to the cuts 
presented in eq.~\eqref{eq:cuts4} (see ref.~\cite{Baur:2003gp}), 
Table~\ref{tab:SMandBSMyields} shows the event yields after applying cuts. 
The background is generated with the cuts of eq.~\eqref{eq:cuts} excluding potential soft and collinear singularities. 
The invariant mass cuts of eq.~\eqref{eq:invmasscuts} significantly
reduce the background and retain the
SM and BSM signal. 
$\Delta R(\gamma, \gamma)$ and $\Delta R(b, \gamma)$
must be chosen differently to optimise for SM and BSM. The cuts of eq.~\eqref{eq:cuts4} 
give the best results for SM 
di-Higgs production and are less efficient for large values of $\lambda_{3H}^{BSM}$.
The cuts of eq.~\eqref{eq:cuts4b} are optimised for $\lambda_{3H}^{BSM} = 10 \lambda_{3H}$. 

\begin{table}
\begin{center}
\begin{tabular}{|c|r|r|r|r|r|}
\hline
&&&&&\\[0.1pt]
Cuts & B & $S(\lambda_{3H})$ & $S(10 \times \lambda_{3H})$  & $S(\lambda_{3H})/\sqrt{B}$ & $S(10\times \lambda_{3H})/\sqrt{B}$ \\[2pt]
\hline 
Before cuts & -- & 128 & 2280  & -- & -- \\
\hline
eq. \eqref{eq:cuts} & 166950 & 48 &726	 &0.12	&1.78 \\
\hline
eq. \eqref{eq:invmasscuts} & 395 & 48	& 726	& 2.43	& 36.53 \\
\hline
eq.\eqref{eq:cuts4}     & 26    & 35	& 189	& 6.80	& 36.85\\
eq.\eqref{eq:cuts4a}    & 41    & 39	& 256	& 6.10	& 40.16 \\
eq.\eqref{eq:cuts4b}    & 240   & 48	& 679	& 3.08	& 43.84 \\
\hline
\end{tabular}
\caption{The number of events in: $b\bar{b}\gamma\gamma$ background, SM and BSM with $\lambda_{3H}^{BSM}=10 \lambda_{3H}$ 
predicted for the luminosity upgraded LHC ($\mathcal{L}=3000$fb$^{-1}$). }
\label{tab:SMandBSMyields}
\end{center}
\end{table}
\section{Conclusions}\label{sec:conclusions}

In view of the upgrade of the LHC to reach higher luminosities
it is vital to study the production and  observability of pairs of Higgs bosons and
how to  measure their self-couplings.
We investigated the trilinear coupling in Higgs pair production in gluon fusion 
with the Higgs bosons decaying into $b\bar{b}\gamma\gamma$. 
With the LHC data collected at 8 TeV ($\mathcal{L}\simeq 20$ fb$^{-1}$)
we envisage 0.5 signal events (Table~\ref{tab:yields}) before experimental reconstruction. 
Recently, the ATLAS collaboration presented the results of a search for Higgs pairs in their dataset~\cite{Aad:2014yja}. 
They observe an excess of di-Higgs candidates with
a significance of $\sim 2\sigma$. 

As the di-Higgs cross-section is small, detailed knowledge on the kinematics is required to improve the separation
between signal and irreducible background. 
Several cross-section predictions in the literature  exploit EFT approximation 
at leading~\cite{Alwall:2011uj} or higher orders~\cite{Dawson:1998py, Baglio:2012np, deFlorian:2013jea}. 
We confronted EFT  with an exact matrix element calculation at leading order.
At $\sqrt{\hat{s}}\simeq 2 m_q$ we find large  differences. 
EFT neglects part of the complicated structure of the box MES which,
adds significantly to the discrepancy due to the triangle contribution.
For the exact calculation we obtain a larger longitudinal boost for and an on average smaller opening angle
between the Higgs bosons,
in agreement with previous studies~\cite{Dawson:2012mk, Li:2013flc, Grigo:2013rya}.

To get a better understanding of how the triangle and box
contribute to the MES, we studied their
behaviour over a large interval of $\sqrt{\hat{s}}$.
We show that the phase space region in which the relative contribution of the self-coupling
maximises is around  $\sqrt{\hat{s}}\simeq m_H$. The contribution to the MES in this region
is, however, very small. 
At large values of $\sqrt{\hat{s}}$ the trilinear coupling plays no role anymore.
This is in agreement with the observations presented in~\cite{Dolan:2012rv}.
For a non-negligible contribution from the trilinear Higgs coupling, experimental searches
should focus on the region where $\sqrt{\hat{s}}< 400$ GeV.

We modified the kinematical cuts proposed 
in ref.~\cite{Baur:2003gp} to study the $b\bar{b}\gamma\gamma$ final state.
They provide a good separation
between di-Higgs signal and irreducible background.
To determine if additional selection criteria should improve the $\lambda_{3H}$ measurement,
we increased $\lambda_{3H}$ by a factor of 10.
This leads to the cuts in~\eqref{eq:cuts4b}. 


\acknowledgments
We would like to thank Michael Spira and Marco Zaro for help in using their programs. M.S. acknowledges a useful
discussion with Tilman Plehn.



\end{document}